\documentclass[12pt]{iopart}
\usepackage{graphicx}
\usepackage{xfrac}
\usepackage{bm}        
\usepackage{amssymb}   
\usepackage{color}
\usepackage{subfig}
\usepackage{mathrsfs}
\usepackage{hyperref}
\usepackage{cases}
\usepackage{latexsym}

\begin{document}

\title[Heading]{Atypical Behavior of Collective Modes in Two-Dimensional Fermi Liquids}

\author{M P Gochan$^1$, J T Heath$^1$, K S Bedell$^1$}

\address{$^1$ Department of Physics, Boston College, Chestnut Hill, MA 02467, United States of America }
\ead{gochan@bc.edu}
\vspace{10pt}
\begin{indented}
\item[]December 2019
\end{indented}

\begin{abstract}
Using the Landau kinetic equation to study the non-equilibrium behavior of interacting Fermi systems is one of the crowning achievements of Landau's Fermi liquid theory. While thorough study of transport modes has been done for standard three-dimensional Fermi liquids, an equally in-depth analysis for two dimensional Fermi liquids is lacking. In applying the Landau kinetic equation (LKE) to a two-dimensional Fermi liquid, we obtain unconventional behavior of the zero sound mode $c_0$. As a function of the usual dimensionless parameter $s=\omega/qv_F$, we find two peculiar results: First, for $|s|>1$ we see the propagation of an undamped mode for weakly interacting systems. This differs from the three dimensional case where an undamped mode only propagates for repulsive interactions and the mode experiences Landau damping for any arbitrary attractive interaction. Second, we find that regardless of interaction strength, a propagating mode is forbidden for $|s|<1$. This is profoundly different from the three-dimensional case where a mode can propagate, albeit damped. In addition, we present a revised Pomeranchuk instability condition for a two-dimensional Fermi liquid as well as equations of motion for the fluid that follow directly from the LKE. In two dimensions, we find a constant minimum for all Landau parameters for $\ell\geq 1$ which differs from the three dimensional case. Finally we discuss the effect of a Coulomb interaction on the system resulting in the plasmon frequency $\omega_p$ exhibiting a crossover to the zero sound mode.     
\end{abstract}

%
%
%
%
%

\section{Introduction}
Since its advent in the late 1950's, Fermi liquid theory \cite{Baym_Pethick,Pines} has become a crucial tool for condensed matter physicists trying to various systems of interacting fermions. Although Fermi liquid theory has many advantages, one of the most significant being the Landau kinetic equation (LKE), which gives theorists a straightforward means to understand non-equilibrium phenomena in a large class of fermionic liquids \cite{Landau1,Landau2,Landau3,Landau4}. Of particular interest is the prediction of zero sound (also known as quantum sound or collisionless sound) that arises from quasiparticle interactions. In a charged Fermi liquid, this mode is replaced by a plasmon mode which exhibits a crossover from a plasmon to zero sound. Zero sound has undergone thorough study from both a theoretical \cite{Gorkov,Mermin,Pethick} and experimental perspective \cite{Abel,Keen}, adding credibility to Landau's work. 
	
Since the work of Haldane \cite{Haldane_Luttinger_liquid}, it has become well-known that the Landau quasiparticle paradigm breaks down in 1D, yet the status of a stable Fermi liquid in a two-dimensional system remains somewhat unclear. Many earlier works on 2D Fermi liquid theory took the Landau-Fermi phenomenology for granted, describing the electric response \cite{Stern, Rajagopal,Toyoda2} and finite-temperature effects \cite{Toyoda,Kim_Coffey} of 2D metals by taking a Fermi-liquid approach or assuming a conventional Fermi distribution for collective excitations in the ground state. More recently, beginning with the pioneering work of Anderson \cite{Anderson1,Anderson2}, higher-dimensional extensions of the celebrated bosonization technique have shown evidence of possible singularities in the Landau parameters as the result of forward scattering \cite{Stamp1,Stamp2}, suggesting that the 2D fermionic liquid would behave more akin to a Luttinger liquid as opposed to a traditional Landau-Fermi liquid. Although studies have shown that the 2D Fermi liquid is stable for relatively weak interactions \cite{Halboth,Feldman1,Feldman2,Feldman3}, microscopic considerations of dimensional reduction on the {\it dynamics} of the 2D Fermi liquid phase are still very much active, with bosonization techniques in the spirit of Anderson hinting at strong deviations from the Landau prediction for the zero sound dispersion \cite{Castro1,Castro2,Barci}. Although similar zero sound behavior is seen in the normalized 2D susceptibility by including vertex corrections to the particle-hole bubble \cite{Maslov}, the question of exactly why the bosonization prediction (which only remains valid in the local theory \cite{Castro1}) disagrees with the traditional result of Landau (which should be valid for dimensions $D\ge 2$ for all partial-wave channels $\ell\ge 0$) remains unresolved. The lack of discussion on Fermi liquid transport in two-dimensional systems within the framework of Landau-Fermi liquid approach, the assumption that three-dimensional expressions are valid in two dimensions without providing proof \cite{Grimmer}, and the recent advancements on unconventional hydrodynamics in two-dimensional fermionic systems  \cite{Lucas, Crossover, Shear} are the main motivations for this work. 
We address these issues by describing the zero sound in a 2D Fermi liquid by incorporating the effects of dimensional phase-space reduction directly onto a partial wave-expansion of the fermionic LKE. Unlike previous attempts, such a calculation naturally follows from Landau's original work on the topic \cite{Landau3, Landau4}, incorporating mathematical extensions of well-known orthogonal functions \cite{Heath} into the hydrodynamic description of collective excitations and subsequently avoiding complications from higher-dimensional bosonization.

This paper is organized as follows: In section 2, we determine the stability conditions of a Fermi liquid in two dimensions in the same way Pomeranchuk did for three dimensional Fermi liquid \cite{Pomeranchuk}. Specific moments of the LKE are derived to show how the stability condition manifests itself in transport phenomena. In section 3, we use the LKE to derive response functions and subsequently analyze the zero sound and plasmon mode in a two-dimensional Fermi liquid. Our response functions differ from those derived by Stern \cite{Stern} and Giuliani \& Vignale \cite{Giuliani} via the inclusion of Landau (interaction) parameters. Additionally, we include higher order Landau parameters to determine if they have a prominent effect on the mode. We close in section 4 by summarizing our results and discussing future work. 

\section{Stability of the two-dimensional Fermi Liquid}
Before applying the LKE to a two-dimensional Fermi liquid, we must first determine the conditions for which the Fermi liquid state is stable in two dimensions. Generally, the ground state energy being stationary and a minimum is the main requirement for (thermodyamic) stability. Originally derived by Pomeranchuk, the constraint on Landau parameters $F_{\ell}^{s,a}$ for stability in a three-dimensional Fermi liquid has been derived multiple times \cite{Baym_Pethick,Pines,Pomeranchuk,Tripathi} and found to be 
\begin{equation}
F_{\ell}^{s,a}>-\left(2l+1\right)
\label{eq:3dStability}
\end{equation}
If the above inequality is violated, it simple means that the interaction is ``too strong" which leads to an instability of the normal state \cite{Pines}. An interesting feature of eqn.(\ref{eq:3dStability}) is how the minimum value for each Landau parameter depends on the specific moment. This is an agreement with the general perturbative nature of Fermi liquid theory; higher order interaction terms must be smaller than previous terms for validity of the theory. 

To derive the stability condition for a two-dimensional Fermi liquid, we start with the change in free energy density $F=E-\mu n$ 
\begin{equation}
\delta F=\frac{1}{V}\sum_{\mathbf{p}\sigma}\left(\varepsilon_{\mathbf{p}\sigma}-\mu\right)\delta n_{\mathbf{p}\sigma}+\frac{1}{2V^2}\sum_{\mathbf{p}\sigma,\mathbf{p}'\sigma'}f_{\mathbf{p}\mathbf{p}'}^{\sigma\sigma '}\delta n_{\mathbf{p}\sigma}\delta n_{\mathbf{p}'\sigma'}
\label{eq:free_energy1}
\end{equation}
Our goal is to obtain an expression for $\delta F$ in terms of Landau parameters and then impose the stability condition. At $T=0$, we linearize the $\left(\varepsilon_{\mathbf{p}\sigma}-\mu\right)$ term and Taylor expand the Heaviside step functions that appear in $\delta n_{\mathbf{p}\sigma}$. For a two-dimensional system, the orthogonal functions used must be Chebyshev polynomials $T_{\ell}(x)$ which are defined as follows \cite{GR}
$$
T_{\ell}(\cos\theta)=\cos\left(\ell\theta\right)
$$
$$
\int_0^{2\pi}d\Omega T_{\ell}(\cos\theta)T_{\ell'}(\cos\theta)=\pi\left(1+\delta_{\ell',0}\right)\delta_{\ell,\ell'}
$$
where $d\Omega$ is the 2d polar angle. The remaining calculation follows from the literature \cite{Heath} and leads to 
\begin{equation}
\delta F=\frac{\pi v_Fp_F}{\left(2\pi\hbar\right)^2}\sum_{\ell}|\nu_{\ell}|^2\left(1+\delta_{\ell,0}+\frac{1}{2}F_{\ell}^{s,a}\left(1+\delta_{\ell,0}\right)^2\right)
\label{eq:free_energy2}
\end{equation}
Imposing that eqn.(\ref{eq:free_energy2}) be greater than zero, we find the condition for stability in a two-dimensional Fermi liquid is given by the following constraint:
\begin{equation}
F_{\ell}^{s,a}>-\frac{2}{1+\delta_{\ell,0}}
\label{eq:2dStability}
\end{equation}
The $\ell=0$ Landau parameter obeys the same stability condition as in the three-dimensional Fermi liquid. A stark difference is seen in higher order terms. In a two-dimensional Fermi liquid, the minimum value Landau parameters for $\ell\geq 1$ is a constant, i.e. $F_{\ell}^{s,a}>-2$ for $\ell\geq 1$. The stability condition is far more strict for a two dimensional Fermi liquid than for its three-dimensional counterpart. This change, a seemingly ``constant" constraint on higher order Landau parameters, can be interpreted in a number of ways. Historically, it's been known that Fermi liquid runs into trouble as dimensionality is reduced. The failure of Fermi liquid theory in one-dimensional interacting Fermi systems is well studied \cite{Solyom,Voit,Haldane} where the lowest energy excitation being a collective boson (as opposed to a traditional Landau quasiparticle), and the gap in the particle-hole spectrum at low energies, invalidates Fermi liquid theory and Luttinger liquid theory must be used. In two-dimensional systems, the distinction between a good and bad Fermi liquid requires more care. While a sharp Fermi surface exists and the discontinuity in the distribution function at the Fermi momentum remains, a two-dimensional Fermi liquid theory is not as robust as it is in three-dimensions. For example, the earlier work of Anderson \cite{Anderson1,Anderson2} discussed the breakdown of Fermi liquid theory, albeit in two-dimensional Hubbard systems and under specific conditions. More recent work on two-dimensional interacting Fermi systems have alluded to possible marginal Fermi liquid behavior \cite{Varma,MarginalFL,DasSarma} where the low temperature behavior of the self energy is modified. These past results are in line with our derived stability condition, eqn.(\ref{eq:2dStability}), which seems to imply the two-dimensional Fermi liquid is not as robust and more susceptible to instabilities.

This stability condition can be readily observed if we derive specific moments of the LKE. The $\ell=0$ moment leads to the number conservation law
\begin{equation}
s\nu_0=\frac{\nu_1}{3}\left(1+\frac{1}{2}F_1^s\right)
\label{eq:NumberConservation}
\end{equation}
while the $\ell=1$ moment leads to the equation of motion for fluid momentum 
\begin{equation}
s\nu_1-\frac{3}{2}\left(1+F_0^s\right)\nu_0-\frac{7}{10}\left(1+\frac{1}{2}F_2^s\right)\nu_2=U
\label{eq:EOM1}
\end{equation}
Eliminating $\nu_1$ from these equations, we get
\begin{equation}
\left(s^2-\left(\frac{c_1}{v_f}\right)^2\right)\nu_0-\frac{7}{30}\left(1+\frac{1}{2}F_1^s\right)\left(1+\frac{1}{2}F_2^s\right)\nu_2=\left(1+\frac{1}{2}F_1^s\right)\frac{U}{3}
\label{eq:EOM2}
\end{equation}
where the speed of ordinary (first) sound $c_1$ is
\begin{equation}
c_1^2=\frac{v_F^2}{2}\left(1+F_0^s\right)\left(1+\frac{1}{2}F_1^s\right)
\label{eq:FirstSound}
\end{equation}
Unlike the results derived by Baym and Pethick \cite{Baym_Pethick}, the coefficients in front of the Landau parameters $F_{\ell}^s$ for $\ell\geq1$ in eqns.(\ref{eq:EOM1}) and (\ref{eq:EOM2}) are all a constant $\frac{1}{2}$ as opposed to different constants. This, along with the expression for ordinary sound eqn.(\ref{eq:FirstSound}), is in agreement with our derived stability condition eqn.(\ref{eq:2dStability}) which further supports the more restrictive nature of the two-dimensional Fermi liquid.

\section{Collective Modes: Zero Sound \& Plasmons}
In order to study the non-equilibrium phenomena of the two-dimensional Fermi liquid, we begin with the LKE
\begin{equation}
\frac{\partial n_{\mathbf{p}\sigma}}{\partial t}-\{\varepsilon_{\mathbf{p}\sigma},n_{\mathbf{p}\sigma}\}=I
\label{eq:LKE}
\end{equation}
where $\{\cdots,\cdots\}$ are Poisson brackets, $n_{\mathbf{p}\sigma}$ is the quasiparticle distribution function, $\varepsilon_{\mathbf{p}\sigma}$ is the quasiparticle Hamiltonian, and $I$ is the collision integral. We linearize and Fourier transform eqn.(\ref{eq:LKE}) and are left with
\begin{equation}
\left(\omega-\mathbf{q}\cdot\mathbf{v}_{\mathbf{p}\sigma}\right)\delta n_{\mathbf{p}\sigma}\left(\mathbf{q},\omega\right)+\left(\mathbf{q}\cdot\mathbf{v}_{\mathbf{p}\sigma}\right)\frac{\partial n_{\mathbf{p}\sigma}^0}{\partial\varepsilon_{\mathbf{p}\sigma}^0}\delta\varepsilon_{\mathbf{p}\sigma}\left(\mathbf{q},\omega\right)=iI
\label{eq:LLKE}
\end{equation}
By definition, the zero sound mode is a collision-less sound mode (a sound mode at zero temperature) which allows us to set $I=0$. Writing $\delta n_{\mathbf{p}\sigma}$ in terms of Fermi surface distortions $\nu_{\mathbf{p}\sigma}$, we can cast eqn.(\ref{eq:LLKE}) in a familiar form as derived by Baym and Pethick \cite{Baym_Pethick}
\begin{equation}
\nu_{\mathbf{p}\sigma}+\frac{\mathbf{q}\cdot\mathbf{v}_{\mathbf{p}\sigma}}{\omega-\mathbf{q}\cdot\mathbf{v}_{\mathbf{p}\sigma}}\sum_{\mathbf{p}'\sigma '}f_{\mathbf{p}\mathbf{p}'}^{\sigma\sigma '}\left(\frac{\partial n_{\mathbf{p}'\sigma'}^0}{\partial\varepsilon_{\mathbf{p}'\sigma'}^0}\right)\nu_{\mathbf{p}'\sigma'}=\frac{\mathbf{q}\cdot\mathbf{v}_{\mathbf{p}\sigma}}{\omega-\mathbf{q}\cdot\mathbf{v}_{\mathbf{p}\sigma}}U
\label{eq:LLKE2}
\end{equation}
It is at this point that we begin to see behavior unique to two-dimensional systems. 

As was done in the previous section, we use the Chebyshev polynomials of first kind where the angle between $\mathbf{p}$ and $\mathbf{p}'$, $\theta$, is planar restricted. The orthogonality relation:
$$
\int_{-1}^1\frac{dx}{\pi}\frac{T_{\ell}(x)T_{\ell'}(x)}{\sqrt{1-x^2}} = \left\{
\begin{array}{ll}
\frac{1}{2}\delta_{\ell\ell'}, & \ell\neq0,\ell'\neq0 \\
\\
1, & \ell=\ell'=0
\end{array} \right.
$$
is what leads to the unusual behavior of the collective mode. Evaluating the second term on the l.h.s. of eqn.(\ref{eq:LLKE2}) leaves us with
\begin{equation}
\sum_{\ell}T_{\ell}(x)\nu_{\ell}+\frac{x}{s-x}\sum_{\ell'}\left(-\frac{1}{2}F_{\ell'}^s\right)\left(1+\delta_{\ell',0}\right)T_{\ell'}(x)\nu_{\ell'}=\frac{x}{s-x}U
\label{eq:Final_LKE}
\end{equation}
where only the spin symmetric Landau parameters matter. The goal is to determine the poles of the response functions $\nu_{\ell}/U$ as these are the frequencies of oscillation of the propagating mode. 

So far, with the exception of introducing Chebyshev polynomials as our orthogonal function basis, this analysis is what's done  in three-dimensional Fermi liquids. A profound difference is in solving for the $l=0$ response function. If we multiply eqn.(\ref{eq:Final_LKE}) by $T_0(x)/\sqrt{1-x^2}$ and exploit the orthogonality relation, we arrive at
\begin{equation}
\nu_0-F_0^s\nu_0\int_{-1}^1\frac{dx}{\pi}\frac{x}{s-x}\frac{1}{\sqrt{1-x^2}} = U\int_{-1}^1\frac{dx}{\pi}\frac{x}{s-x}\frac{1}{\sqrt{1-x^2}}
\label{eq:zero_sound_LKE}
\end{equation}
where we've imposed $F_{\ell'}^s=0$ for $\ell'\geq 1$. The integrals in eqn.(\ref{eq:zero_sound_LKE}) are highly non-trivial and need to be evaluated carefully. Details of the calculation can be found in \cite{Heath}. The difficulty arises from the integration bounds also being singularities of the integrand. To circumvent this issue, a careful change of variables must be made which turns the above integrals into
$$
\int_{-1}^1\frac{x}{s-x}\frac{1}{\sqrt{1-x^2}}dx=\pm\frac{1}{2i}\oint_{|z|=1}\frac{(z-i)(z+i)}{z(z-s-\sqrt{s^2-1})(z-s+\sqrt{s^2-1})}
$$
where we assume a general complex form for $s$: $s=s_1+is_2$ and the contour is the unit circle. The integral is then evaluated for $|s|>1$ and $|s|<1$ leading to
\begin{equation}
\int_{-1}^1\frac{dx}{\pi}\frac{x}{s-x}\frac{1}{\sqrt{1-x^2}}=-\left(1-\frac{|s|}{\sqrt{s^2-1}}F(s)\right)
\label{eq:Integral}
\end{equation}
where the function $F(s)$ is defined as
$$
F(s)=1-\left[\delta(s_1)\Theta\left(|s|-1\right)+\delta(s_2)\Theta\left(1-|s|\right)\right]
$$
and captures the behavior of $|s|>1$ and $|s|<1$ which determines whether the poles of the integral are inside/outside the contour. Finally, the response function $\nu_0/U$, and subsequently the zero sound mode, can then be determined depending on the value of $|s|$. 

\subsection{Response function for $|s|>1$}
\begin{figure}[h]
	\centering
	\includegraphics[scale=0.75]{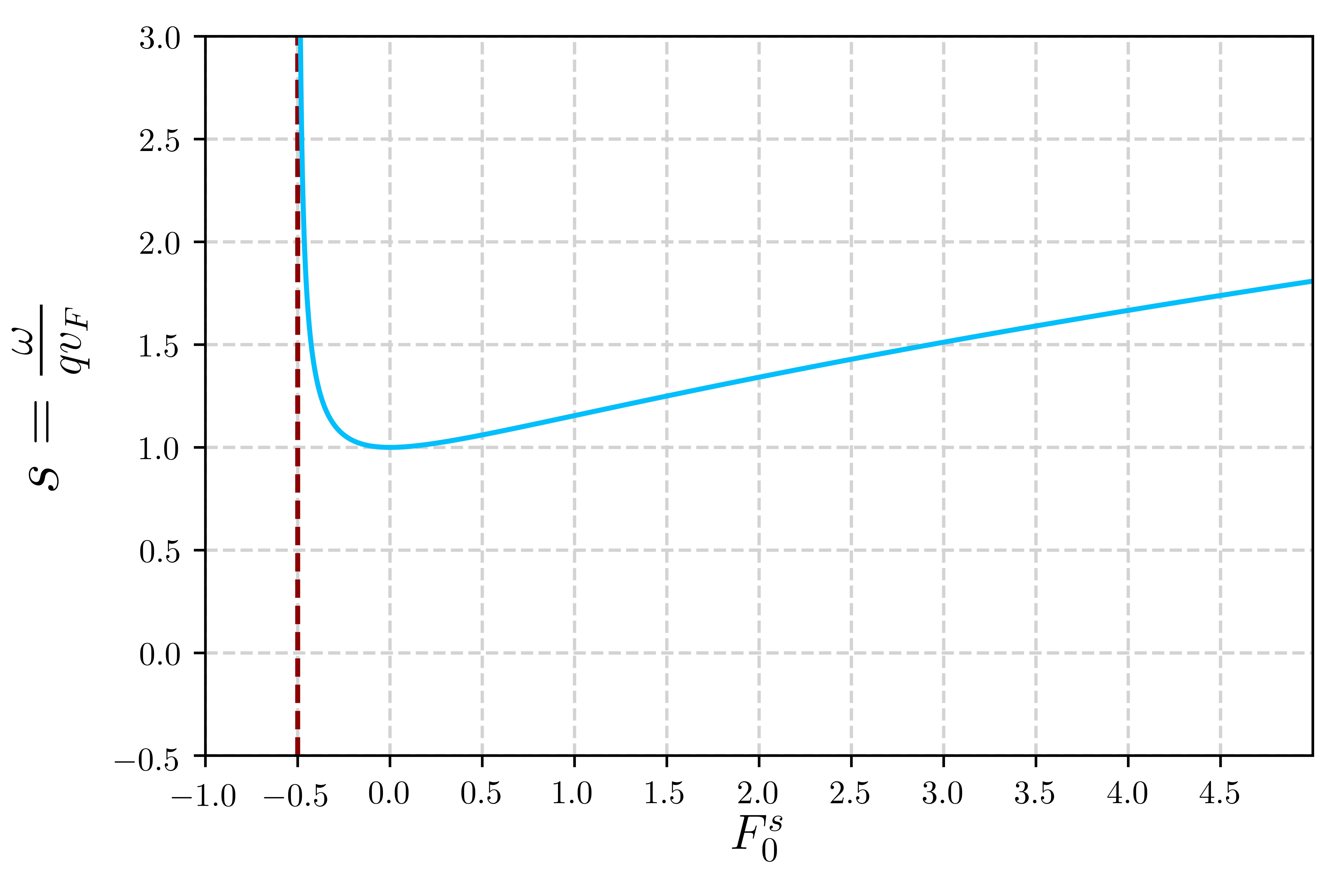}
	\caption{(COLOR ONLINE) A plot of $s$ as a function of $F_0^s$ for $|s|>1$ (blue curve) given by eqn.(\ref{eq:zerosound1}). The dashed red line at $F_0^s=-0.5$ indicates a dynamic instability, different from the instability discussed in the previous section, of the zero sound mode. The unusual behavior of this mode can be seen for attractive interactions $-0.5<F_0^s<0$ where undamped propagation occurs. This is in stark contrast to the behavior found in three-dimensional Fermi liquids where, for arbitrarily small attractive interactions, the presence of Landau damping strongly inhibits the mode \cite{Baym_Pethick}. Although experimental realization of a two-dimensional Fermi liquid with attractive interactions is due to the work of Fr\"ohlich et. al. \cite{Frohlich}, to our knowledge, a theoretical description of the two-dimensional Fermi liquid with attractive interaction has not been carried out.}
	\label{fig:ZeroSound1}
\end{figure}

For values of $|s|>1$, the response function
\begin{subnumcases}{\frac{\nu_0}{U}=}
	-\left(\left(1-\frac{s}{\sqrt{s^2-1}}\right)^{-1}+F_0^s\right)^{-1} & $s_1\neq 0$  
	\label{eq:Response1a}
	\\
	-\left(1+F_0^s\right)^{-1} & $s_1=0$
	\label{eq:Response1b}
\end{subnumcases}
First focusing on the response function when the real part of $s$ is non-zero, given in eqn.(\ref{eq:Response1a}), and solving for the poles, we get the following two equations
\begin{subequations}
	\begin{equation}
		\frac{s_2\sqrt{s^2-1}}{2s^2-2s_1\sqrt{s^2-1}-1}=0
		\label{eq:Response1c}
	\end{equation}
	\begin{equation}
		\frac{s^2-s_1\sqrt{s^2-1}-1}{2s^2-2s_1\sqrt{s^2-1}-1}=-F_0^s
		\label{eq:Response1d}
	\end{equation}
\end{subequations} 
Eqn.(\ref{eq:Response1c}) immediately sets the imaginary part of $s$ to be zero (i.e. $s_2=0$) and subsequently sets $s$ to be purely real (i.e. $s=s_1$). Solving eqn.(\ref{eq:Response1d}) for $s$ yields the following for the propagating zero sound mode in a two-dimensional Fermi liquid
\begin{equation}
s=\frac{\omega}{qv_F}=\frac{1+F_0^s}{\sqrt{1+2F_0^s}}
\label{eq:zerosound1}
\end{equation}
which is plotted in fig.(\ref{fig:ZeroSound1})\footnote{The negative solution of eqn.(\ref{eq:Response1d}) is disregarded as this mode can't be excited at $T=0$}. It is worth noting that the result given above is identical to the 2D zero sound dispersion found by taking a simple Fourier expansion of the bosonized field near the 2D system's Fermi surface \cite{Castro1}. However, as we will soon see (and as we have already seen in the case of the 2D Pomeranchuk instability), the inclusion of the full Chebyshev orthogonality condition in the expansion of the Landau parameter will have drastic results on the zero sound dispersion unaccounted for in the contemporary literature. 

\begin{figure}[h]
	\centering
	\includegraphics[scale=0.75]{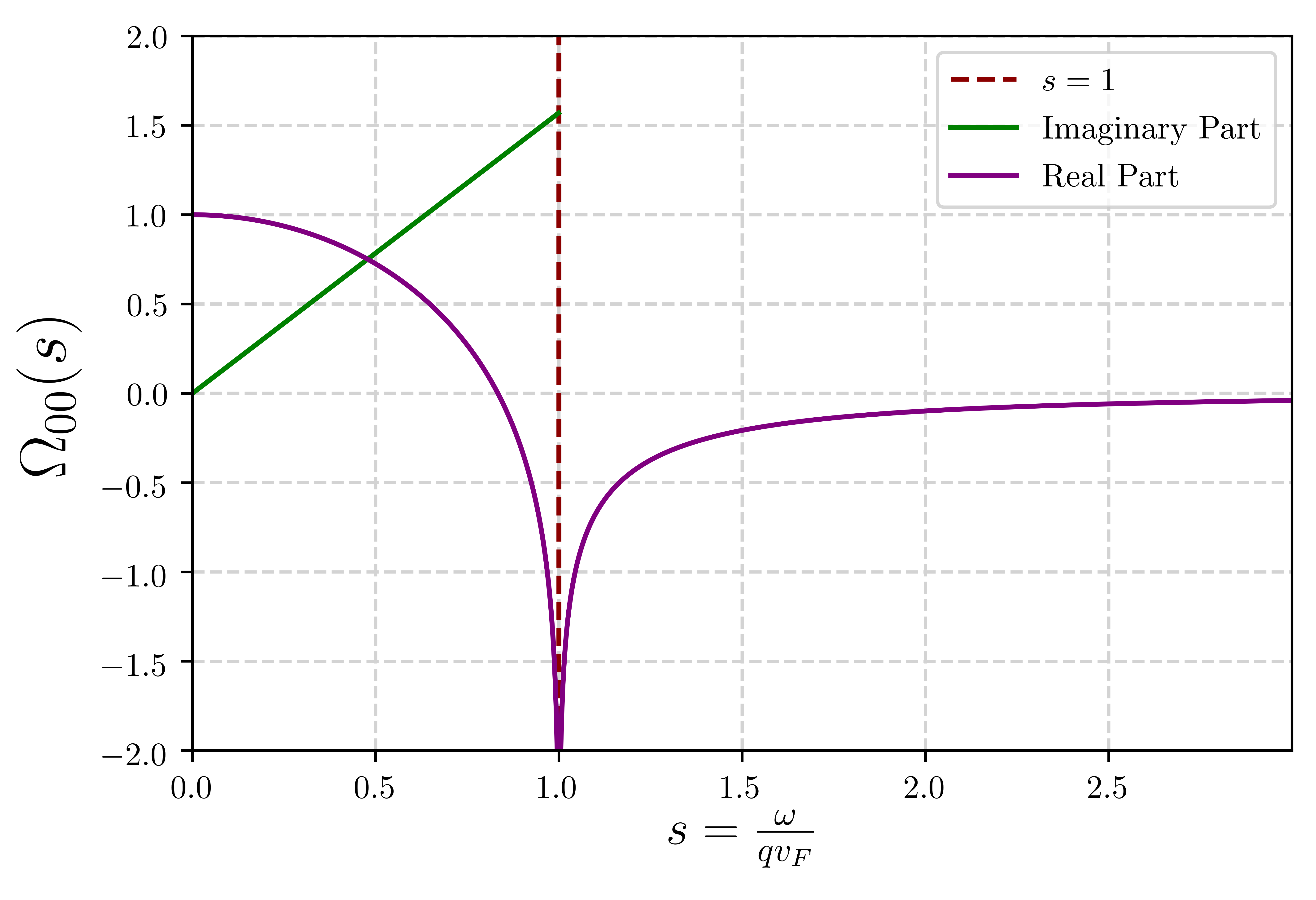}
	\caption{(COLOR ONLINE) A plot of $\Omega_{00}$ as a function of $s$ given by eqn.(1.3.16) from Baym and Pethick \cite{Baym_Pethick}. In solving for the zero sound mode in three dimensions, one solves $1+F_0^s\Omega_{00}=0$ for $s$ after imposing different limits on $F_0^s$. If $F_0^s>0$, we see an undamped zero sound mode and if $F_0^s<0$, we see a Landau damped mode. This is very different from the two dimensional case where an undamped propagating mode can exist for an attractive interaction, and Landau damping is seemingly absent as $s$ is either purely real or purely imaginary.}
	\label{fig:3D}
\end{figure}

If we consider $F_0^s\gg1$, we see 
$$
s\approx\sqrt{\frac{F_0^s}{2}}\rightarrow c_0\approx v_F\sqrt{\frac{F_0^s}{2}}
$$
while for $F_0^s\ll1$,
$$
s\approx1+\frac{1}{2}\left(F_0^s\right)^2\rightarrow c_0\approx v_F\left(1+\frac{1}{2}\left(F_0^s\right)^2\right)
$$
both of which can be seen in the behavior of the blue curve in fig.(\ref{fig:ZeroSound1}). The interesting behavior of the two-dimensional zero sound mode is seen if we consider attractive interactions (i.e. $F_0^s<0$). As per Baym \& Pethick \cite{Baym_Pethick}, for arbitrarily small attractive interactions, Landau damping strongly affects the zero sound mode. However, as discussed in previous works \cite{Pines,Sergio,Yukawa}, a non-zero real and imaginary part of the response function is crucial to the existence of Landau damping. This is easily seen in the response function for a three-dimensional Fermi liquid as shown in fig.(\ref{fig:3D}). As we've shown earlier, for $|s|>1$ and non-zero real part, the imaginary part must be zero. Thus the mode can propagate freely for a certain range of attractive interactions $-0.5<F_0^s<0$. 

Turning our attention to the response function given by eqn.(\ref{eq:Response1b}), we see that $\nu_0/U$ is independent of $s$. This response function violates causality, established through the well-known Kramers-Kronig relations \cite{Pines}. As such, we disregard this solution and conclude that $s$ must be purely real for $|s|>1$.

\subsection{Response function for $|s|<1$}
\begin{figure}[h]
	\centering
	\includegraphics[scale=0.75]{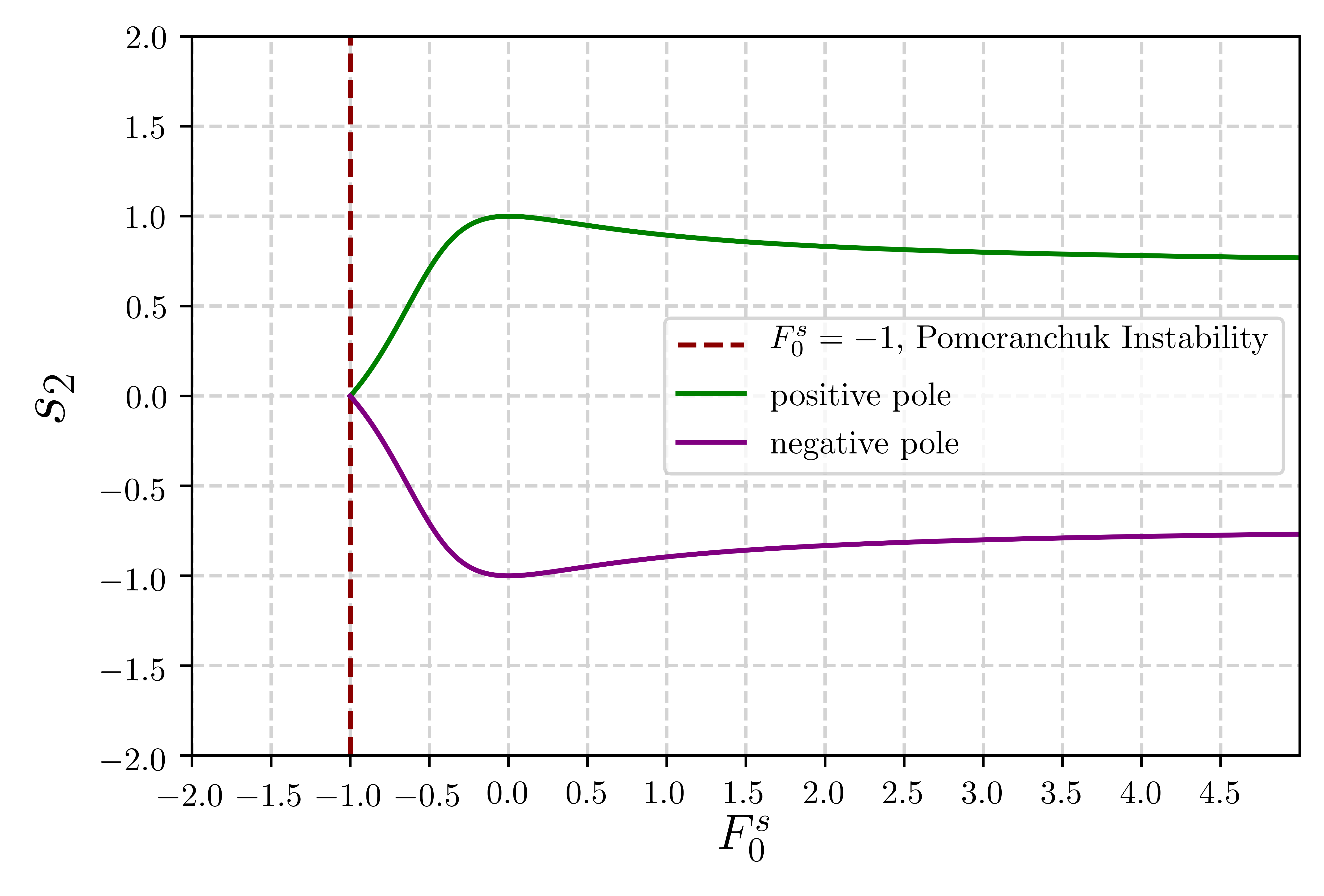}
	\caption{(COLOR ONLINE) A plot of $s_2$ as a function of $F_0^s$ for $|s|<1$ given by eqn.(\ref{eq:zerosound2}). The dashed red line at $F_0^s=-1$ indicates the Pomeranchuk instability as derived in the previous section. While both poles seem to present possible modes for $|s|<1$, they're both invalid for different reasons. The positive pole (green curve) is located in the upper half of the complex plane and therefore violates the well-known Kramer's Kronig relations. The negative pole (purple curve) is purely imaginary (as is $s$ in general for $|s|<1$) indicating a severely damped mode that cannot propagate. This is different from Landau damping, which requires both nonzero real and imaginary parts to $s$ \cite{Pines,Sergio,Yukawa}.}
	\label{fig:sLessThanOne}
\end{figure}

For values of $|s|<1$, we have
\begin{subnumcases}{\frac{\nu_0}{U}=}
-\left(\left(1-\frac{s}{\sqrt{s^2-1}}\right)^{-1}+F_0^s\right)^{-1} & $s_2\neq 0$  
\label{eq:Response2a}
\\
-\left(1+F_0^s\right)^{-1} & $s_2=0$
\label{eq:Response2b}
\end{subnumcases}
First focusing on the situation given in eqn.(\ref{eq:Response2a}), and solving for the poles, we get the following two equations
\begin{subequations}
	\begin{equation}
	\frac{-s_1\sqrt{1-s^2}}{1-2s_2\sqrt{1-s^2}}=0
	\label{eq:Response2c}
	\end{equation}
	\begin{equation}
	\frac{1-s^2-s_2\sqrt{1-s^2}}{1-2s_2\sqrt{1-s^2}}=-F_0^s
	\label{eq:Response2d}
	\end{equation}
\end{subequations} 
Eqn.(\ref{eq:Response2c}) immediately sets the real part of $s$ to be zero (i.e. $s_1=0$) and subsequently sets $s$ to be purely imaginary (i.e. $s=is_2$). Solving eqn.(\ref{eq:Response2d}) for $s$ yields the following for zero sound mode in a two-dimensional Fermi liquid if $|s|<1$
\begin{equation}
s=\pm i\left(\frac{1+F_0^s}{\sqrt{1+2F_0^s+2\left(F_0^s\right)^2}}\right)
\label{eq:zerosound2}
\end{equation}
which is plotted in fig.(\ref{fig:sLessThanOne}). 

Discussion of the results plotted in fig.(\ref{fig:sLessThanOne}) requires more care than the discussion of the results in fig.(\ref{fig:ZeroSound1}). First note that in eqn.(\ref{eq:zerosound2}), we have kept both positive and negative solutions as the earlier argument in regards to negative solutions and $T=0$ only holds assuming we're dealing with real frequencies. Analyzing the positive pole (green curve) first, we see an immediate violation of causality as this pole is in the upper complex plane. The remaining negative pole is purely imaginary and indicates a severely damped mode (i.e. no propagation allowed) that is distinctly different from the usual Landau damping due to the absence of a non-zero real part. Focusing on the alternative solution for the response function given by eqn.(\ref{eq:Response2b}), we see another response function constant in $s$. As per before, this result violates causality and is disregarded. Therefore we conclude that no mode can propagate for values of $|s|<1$.

As we've seen from the above analysis, behavior of the zero sound mode in a two-dimensional Fermi liquid is drastically different from that in its three-dimensional counterpart. Previous work on zero sound in three-dimensions is numerous and all shows the same usual behavior: undamped propagation for $F_0^s>0$ and/or $|s|>1$ (outside the particle-hole continuum), and Landau damping when $F_0^s<0$ and/or $|s|<1$ (inside the particle-hole continuum). While our work does show an undamped mode for repulsive interactions and $|s|>1$, we see propagation for a small region of attractive interaction which is absent in three-dimensions; a sound mode may propagate in the presence of attractive interactions but this is first sound, not zero sound. Additionally, it appears that Landau damping is absent and a Landau damped mode is completely damped out in a two-dimensional Fermi liquid. 

\subsection{Including higher order Landau parameters}
We stress that the above calculation was performed for $\ell=0$ distortions of the Fermi surface and assuming all Landau parameters essentially zero except $F_0^s$. A natural question to ask is if including higher order Landau parameters affects the behavior of the zero sound mode. Specifically, we want to ask whether or not Landau damping can be restored upon inclusion of the $F_1^s$ Landau parameter. 

Going back to eqn.(\ref{eq:Final_LKE}), we multiply by $T_1(x)/\sqrt{1-x^2}$ and exploit the orthogonality relation, arriving at an equation similar to eqn.(\ref{eq:zero_sound_LKE}) with integrals of the following form
$$
\int_{-1}^1\frac{x^n}{s-x}\frac{1}{\sqrt{1-x^2}}dx
$$
Evaluation of the integral above leads to a generalization of the integral in eqn.(\ref{eq:Integral}). Their result allows us to express eqn.(\ref{eq:zero_sound_LKE}) in the following general form, similar to what is done in \cite{Baym_Pethick} for a three-dimensional Fermi liquid
\begin{equation}
\frac{\pi}{2}\left(1+\delta_{\ell 0}\right)\nu_{\ell}+\sum_{\ell'}Q_{\ell\ell'}\left(1+\delta_{\ell'0}\right)F_{\ell'}^s\nu_{\ell'}=-2Q_{\ell 0}U
\label{eq:2D_LKE}
\end{equation}
where $Q_{\ell\ell'}$, a function of $s$, is defined as 
$$
Q_{\ell\ell'}=Q_{\ell'\ell}=\frac{1}{2}\int_{-1}^1\frac{T_{\ell}(x)T_{\ell'}(x)}{\sqrt{1-x^2}}\frac{x}{x-s}dx
$$

If we wish to include the $F_\ell^s$ term, we take the $\ell=0$ and $\ell=1$ terms of eqn.(\ref{eq:2D_LKE})
\begin{subequations}
	\begin{equation}
	\pi\nu_0+2Q_{00}F_0^s\nu_0+Q_{01}F_1^s\nu_1=-2Q_{00}U
	\label{eq:2D_LKE_1}
	\end{equation}
	\begin{equation}
	\frac{\pi}{2}\nu_1+2Q_{10}F_0^s\nu_0+Q_{11}F_1^s\nu_1=-2Q_{10}U
	\label{eq:2D_LKE_2}
	\end{equation}
\end{subequations} 
where we impose $F_l^s=0$ for $l\geq 2$. Solving for the response function we get
\begin{equation}
\frac{\nu_0}{U}=\frac{-2Q_{00}-2F_1^s\frac{\left(Q_{01}\right)^2}{\frac{\pi}{2}+Q_{11}F_1^s}}{\pi+2F_0^sQ_{00}-2F_0^sF_1^s\frac{\left(Q_{01}\right)^2}{\frac{\pi}{2}+Q_{11}F_1^s}}
\label{eq:2D_Response_F1}
\end{equation}
and subsequently solve for $s$ such that the denominator of eqn.(\ref{eq:2D_Response_F1}) is zero. As before, we obtain modes that freely propagate  ($s$ purely real), modes that violate causality (response function approaching a constant value as $s\rightarrow\infty$), and modes that are severely damped ($s$ purely imaginary). However, a new feature that arises from retaining up to the $F_1^s$ Landau parameter is the possibility of a Landau damped collective mode. For $|s|>1$, we have
\begin{equation}
s=s_1+is_2
\label{eq:Response3a}
\end{equation}
where
\begin{subequations}
	\begin{equation}
	s_1=\frac{1}{2}\sqrt{s^2-1}\left(\frac{F_1^s+1}{F_1^s-2}\right)
	\label{eq:3aReal}
	\end{equation}
	\begin{equation}
	s_2=\pm\frac{1}{2}\sqrt{\frac{\left(2+F_1^s\right)\left(-6+F_0^s\left(F_1^s-2\right)+5F_1^s-\left(F_1^s+2\right)s^2\right)}{F_1^s-2}}
	\label{eq:3aIm}
	\end{equation}
\end{subequations} 
and for $|s|<1$, we have
\begin{equation}
s=s_3+is_4
\label{eq:Response3b}
\end{equation}
where
\begin{subequations}
	\begin{equation}
	s_3=\frac{\left(F_1^s+2\right)\sqrt{1-s^2}}{2\left(F_1^s-2\right)}
	\label{eq:3bReal}
	\end{equation}
	\begin{equation}
	s_4=\frac{F_1^s+2}{2\left(F_1^s-2\right)}\left(\sqrt{1-s^2}\pm\sqrt{\frac{6F_1^s+4F_0^s\left(F_1^s-2\right)-4}{F_1^s+2}-2s^2}\right)
	\label{eq:3bIm}
	\end{equation}
\end{subequations} 
Expressions for $s_1$, $s_2$, $s_3$, and $s_4$ solely as function of $F_0^s$ and $F_1^s$ are straightforward to obtain, however they are cumbersome. The importance of eqns.(\ref{eq:Response3a})-(\ref{eq:3bIm}) is their complex nature which allows for the possibility of Landau damping subject to the following constraints on $F_1^s$:
\begin{subnumcases}{F_1^s>}
\frac{-1+\sqrt{53+F_0^s\left(29+4F_0^s\right)}}{4+F_0^s} & $|s|>1$  
\label{eq:F1Condition1}
\\
\frac{2+4F_0^s}{3+2F_0^s} & $|s|<1$
\label{eq:F1Condition2}
\end{subnumcases}
and subsequently on $|s|$
\begin{subnumcases}{|s|<}
\sqrt{\frac{4F_0^sF_1^s+5F_1^s-8F_0^s-6}{F_1^s+2}} & $|s|>1$  
\label{eq:sCndition1}
\\
\sqrt{\frac{2F_0^sF_1^s+3F_1^s-4F_0^s-2}{F_1^s+2}} & $|s|<1$
\label{eq:sCondition2}
\end{subnumcases}
which are derived from imposing that $|s|$ must be real. 
\begin{figure}[h]
	\centering
	\includegraphics[scale=0.75]{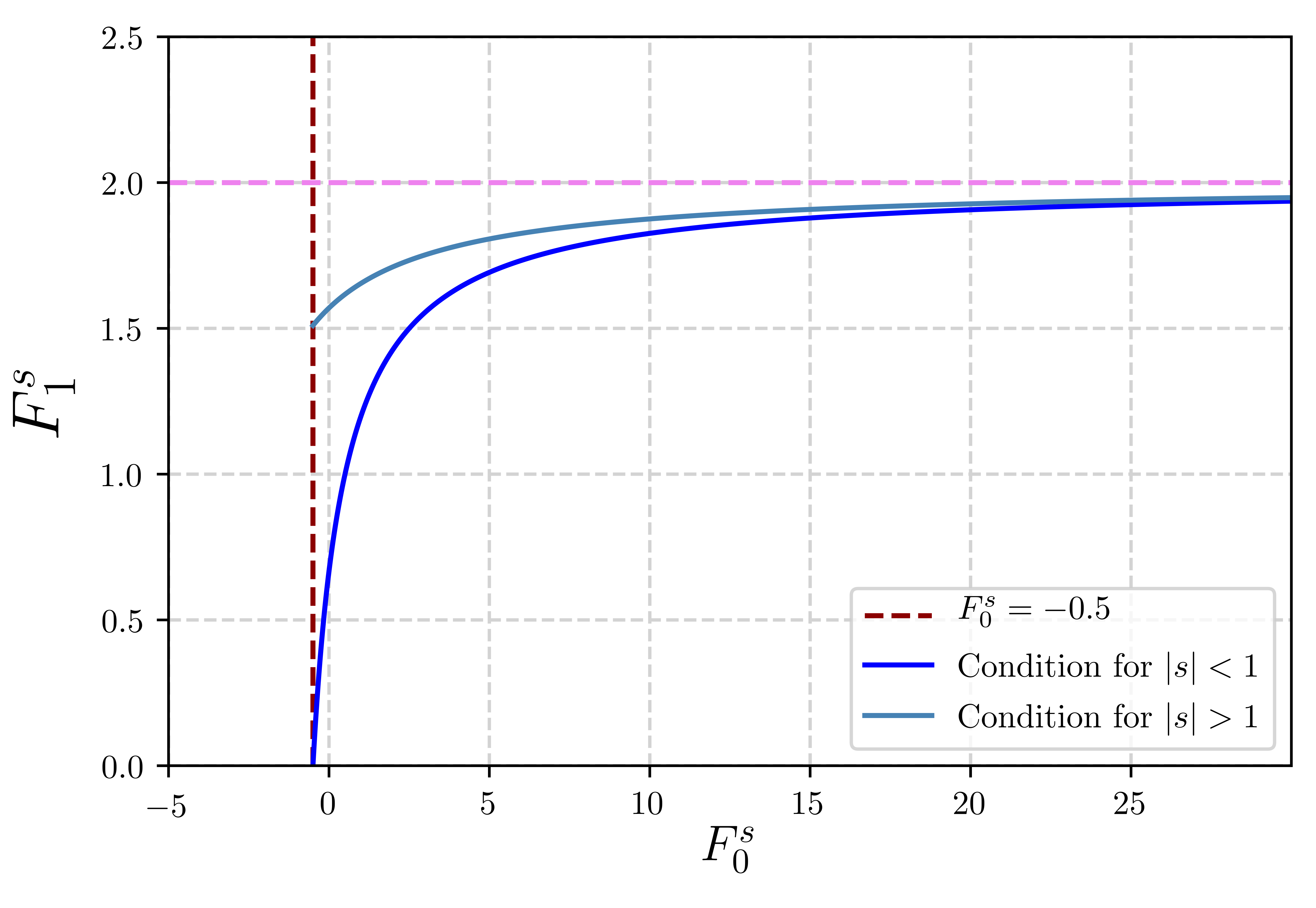}
	\caption{(COLOR ONLINE) Plot of eqn.(\ref{eq:F1Condition1}) (light curve) and eqn.(\ref{eq:F1Condition2}) (dark curve). The curves show the lower bound on $F_1^s$ (as a function of $F_0^s$). For large values of $F_0^s$, we see that the lower bound condition saturates at $F_1^s=2$. For small values of $F_0^s$, the situation is different depending on $|s|$. For $|s|>1$, the smallest value for $F_1^s$ above which Landau damping can occur is $F_1^s=1.5$ while for $|s|<1$, as we approach the dynamic instability at $F_0^s=-0.5$, $F_1^s$ needs to simply be larger than zero for Landau damping to be present. With a large lower bound of $F_1^s\geq 2$ for large values of $F_0^s$, it appears that the phenomena of Landau damping is still absent for $|s|>1$ in spite of including the $F_1^s$ Fermi liquid parameter. For $|s|<1$, Landau damping seems more likely and this agrees with what is seen in the three-dimensional case (The $|s|<1$ mode is inside the particle-hole continuum indicating a Landau damped mode) with the exception of needing the addition of $F_1^s$.}
	\label{fig:F1conditions}
\end{figure}

The above analysis, assuming the next order Landau parameter $F_1^s$ is non-zero, is summarized in fig.(\ref{fig:F1conditions}). As we can see, the inclusion of a non-zero $F_1^s$ term is crucial for the possibility of Landau damping in the system albeit doesn't guarantee it. If $|s|>1$, $F_1^s>1.5$ for small values of $F_0^s$ and $F_1^s>2$ for large values of $F_0^s$. Both scenarios require a rather large value of $F_1^s$ therefore making the presence of Landau damping unlikely. If $|s|<1$, Landau damping is more likely for small values of $F_0^s$ since the lower bound on $F_1^s$ tends to zero as $F_0^s\rightarrow-0.5$. To fully understand Landau damping in a two-dimensional Fermi liquid, one would need values for $F_0^s$ and $F_1^s$ obtained via experiments, for example on the specific heat and compressibility, on two-dimensional interacting Fermi systems. To our knowledge, such data isn't currently available and therefore we're left to make predictions based on the results from above. As such, we conclude that in spite of including an additional Fermi liquid parameter, $F_1^s$, Landau damping is still not present at least for large values of $F_0^s$. For small $F_0^s$, Landau damping remains unlikely for $|s|>1$. If $|s|<1$, the situation resembles that of a three-dimensional Fermi liquid and the mode is immediately inside the particle-hole continuum indicating its damped behavior. 

\subsection{Plasmons}
In practice, most systems under investigation have non-zero charge and as such the zero sound mode discussed throughout this paper needs to be replaced with a plasmon mode. Emergence of plasmon modes within the Landau theory requires inclusion of the Coulomb interaction in eqn.(\ref{eq:LLKE2}). Discussion of the process, including the subtleties when dealing with the singularity at $q=0$ and establishing screening, for a three-dimensional Fermi liquid can be found in Pines \& Nozieres \cite{Pines}. Carrying out the calculation, we arrive at the following for the frequency of a plasmon mode in a three-dimensional Fermi liquid
\begin{equation}
\omega^2=\left(\omega_p^2+\frac{1}{3}q^2v_F^2F_0^s\right)\left(1+\frac{1}{3}F_1^s\right)
\label{eq:3dPlasmon}
\end{equation}
where $\omega_p$ is the RPA plasmon frequency in a  three-dimensional electron gas \cite{DasSarma2}: 
$$
\omega_p=\sqrt{\frac{4\pi ne^2}{m}}
$$
Eqn.(\ref{eq:3dPlasmon}) highlights two new features that arise from using the LKE to analyze a charged Fermi liquid: an effective mass $(F_1^s)$ correction to the plasmon frequency and a crossover to the zero sound mode as function of $q$. 

The general analysis conducted earlier for zero sound doesn't appreciably change when looking at the plasmon mode. The biggest difference comes in altering the Fermi liquid interaction to incorporate the Coulomb interaction 
$$
f_{\mathbf{p}\mathbf{p}'}^s\rightarrow\frac{2\pi e^2}{q}+f_{\mathbf{p}\mathbf{p}'}^s
$$ 
Continuing as before, assuming only $F_0^s$ is non-zero and looking at $s\gg 1$ and $|s|>1$ since the plasmon is a high frequency mode, we arrive at 
\begin{equation}
\omega^2=\frac{1}{1-\frac{1}{4}F_1^s}\left(\omega_p^2+\frac{1}{2}q^2v_F^2F_0^s\right)\left(1+\frac{1}{2}F_1^s\right)
\label{eq:2dPlasmon}
\end{equation}
where now $\omega_p$ is the RPA plasmon frequency in a two-dimensional electron gas. 
$$
\omega_p=\sqrt{\frac{2\pi ne^2q}{m}}
$$
Eqn.(\ref{eq:2dPlasmon}) also exhibits a Fermi liquid correction to $\omega_p$ and an additional term which is the zero sound mode at large $F_0^s$ in a two-dimensional Fermi liquid discussed earlier. 

An interesting difference between eqn.(\ref{eq:3dPlasmon}) and eqn.(\ref{eq:2dPlasmon}) is in the $q$ dependence. In a three-dimensional Fermi liquid, the plasmon mode is robust as it avoids the particle-hole contiuum and remains undamped until $q$ gets very large. However, in a two-dimensional Fermi liquid, the plasmon mode has a linear $q$ dependence and could theoretically be damped from all values of $q$. It appears the undamped propagation of plasmon modes in a two-dimensional Fermi liquid is sensitive to factors such as mass, particle density, $F_0^s$ and $F_1^s$ so as to avoid the particle-hole continuum. As stated earlier, and what can be clearly seen with our discussion of plasmons, experimental data for the Fermi liquid parameters $F_0^s$ and $F_1^s$ is crucial to our understanding of transport phenomena in two-dimensional systems.

\section{Summary}
The work presented here is a preliminary investigation into the transport phenomena in a two-dimensional Fermi liquid using the Landau kinetic equation. With the renewed interest in two-dimensional systems with a possible Fermi liquid state, the results presented here should be of importance as they show profound differences from normal three-dimensional Fermi liquid behavior. The derived stability condition eqn.(\ref{eq:2dStability}) shows that the two-dimensional Fermi liquid is more susceptible to instabilities thus not being as robust as its three-dimensional counterpart. While this point was conjectured in the early years of Fermi liquid theory, a quantitative statement like the one we've shown here was never put forward. 

Where our work differs from those of the past is in our use of the Chebyshev polynomials $T_{\ell}(x)$. The unique form of the weighting function for $T_{\ell}(x)$ led to an abundance of new phenomena emerge. First, focusing on zero sound, we see the possibility of undamped propagation for attractive interactions. This feature is absent in three-dimensional Fermi liquids, where the presence of attractive interactions leads to a Landau damped mode. Second, we see that propagation of zero sound is more restrictive in two-dimensions. Up to first order in the Fermi liquid interactions, zero sound does not propagate for $|s|<1$ and Landau damping is entirely absent unless higher order interaction parameters are considered. Moving onto plasmons, we see an effective mass correction to the normal plasmon frequency and an additional term which is similar in structure to the zero sound mode. This indicates the possibility of a crossover from plasmon to zero sound; similar to a phenomena studied in two-dimensional graphene \cite{Lucas}. Such new behavior is in stark contrast to what happens in a three-dimensional Fermi liquid and should be considered when studying low-dimensional interacting Fermi systems.

The use of Fermi liquid theory to analyze systems may be dated, but it is nonetheless robust and should still be considered in studying interacting Fermi systems. Specifically, our work could be of use in the study of two-dimensional Dirac materials \cite{Dirac,Katsnelson}, such as graphene, when $\mu\neq0$ and the system is in its Fermi liquid state. And although, as it currently stands, the use of LKE at/near the charge neutrality point is invalid, we notice a similarity in the results between our work and that done by Lucas \cite{Lucas} and Svintsov \cite{Crossover} which used other methods to investigate the behavior of collective modes in these exotic systems. Such a similarity seems to suggest that the behavior of electrons near the charge neutrality point might be more Fermi liquid-like than we think and hints at a more robust nature of Fermi liquid theory. 

\section*{Acknowledgements}
The authors would like to thank Mason Gray for useful discussion on transport phenomena in two dimensional systems. This work was partially supported in part by the John H. Rourke Endowment Fund at Boston College.

\section*{References}
\bibliographystyle{iopart-num}
\bibliography{ExoticCollective}

\end{document}